\documentclass[twocolumn,prl,aps,showpacs]{revtex4}
\usepackage{graphicx}

\begin{document}

\title{\bf Thermodynamics of the System of Distinguishable Particles}

\author{Chi-Ho Cheng}
\email{phcch@cc.ncue.edu.tw}  \affiliation{Department of Physics,
National Changhua University of Education, Taiwan }

\date{\today}

\begin{abstract}
The issue of the thermodynamics of a system of distinguishable
particles is discussed in this paper. In constructing the
statistical mechanics of distinguishable particles from the
definition of Boltzmann entropy, it is found that the entropy is
not extensive. The inextensivity leads to the so-called Gibbs
paradox in which the mixing entropy of two identical classical
gases increases. Lots of literature from different points of view
were created to resolve the paradox. In this paper, starting from
the Boltzmann entropy, we present the thermodynamics of the system
of distinguishable particles. A straightforward way to get the
corrected Boltzmann counting is shown. The corrected Boltzmann
counting factor can be justified in classical statistical
mechanics.
\end{abstract}


\maketitle

\vspace{-0.10cm}

\section{Introduction}
The concepts of distinguishable and indistinguishable particles is
important in Statistical Mechanics as their corresponding
entropies are different. The entropy in statistical mechanics is
defined in terms of the logarithm of the number of the accessible
microstates in the phase space. The definition of the entropy is
called Boltzmann entropy in which it is adopted in popular
textbooks \cite{landau, pathria, reif, swendsen}. The microstates
numbers for distinguishable and indistinguishable particles are
certainly different and then their corresponding Boltzmann
entropies are different, too. However, it is not trivial to know
whether the distinguishability property may lead to different
physical results. For example, the Gibbs paradox
\cite{gibbs,sw1,sw2,sw3,re1,re2,re3,re4,re5,re6,re7,re8,re9,re10,re11,re12,re13,re14,re15,re16,re17,re18,re18a,re19,re20,re21}
presents one of the cases.

In its simplest case of the Gibbs paradox, the entropy of two
identical volume of (distinguishable) ideal gas increases after
mixture in which it violates our intuition. Consider two
subsystems of equal volumes $V$ and particle numbers $N$ are
separated by a wall, the Boltzmann entropy of one system is
\begin{eqnarray}
S = k \ln V^N
\end{eqnarray}
Hence the total entropy $S_i$ of two systems is double,
\begin{eqnarray} \label{before}
S_i = 2 k \ln V^N
\end{eqnarray}
in which the additivity of the Boltzmann entropy is assumed. Now
the wall is removed, the gases of two systems mix. The entropy
$S_f$ is
\begin{eqnarray} \label{after}
S_f =  k \ln (2 V)^{2N}
\end{eqnarray}
Thus the mixing entropy reads
\begin{eqnarray}
\Delta S &=& S_f - S_i \nonumber \\
&=& 2 N k \ln 2
\end{eqnarray}
which is positive meaning that the mixing process is irreversible
according to the second law of thermodynamics. Here we also assume
the Boltzmann entropy is equivalent to the usual thermodynamical
entropy (that is, the entropy identified in thermal properties).

The essence of the paradox is in fact that the entropy of the
distinguishable particles is not extensive. The entropy of the
mixture is not equal to the sum of their partition. To resolve the
Gibbs paradox, one introduces the indistinguishable particles in
which a permutation factor $1/N!$ ($N$ is the total particle
number) is included in the total microstate number to overcome the
overcounting \cite{gibbs}. Hence, the entropy before mixing in
Eq.(\ref{before}) should be modified as
\begin{eqnarray}
S_i = 2 k \ln \left( \frac{V^N}{N!} \right)
\end{eqnarray}
and similarly, the entropy after mixing in Eq.(\ref{after}) reads
\begin{eqnarray}
S_f =  k \ln \left( \frac{(2 V)^{2N}}{(2N)!} \right)
\end{eqnarray}
The mixing entropy is
\begin{eqnarray}
\Delta S &=& 2 N k \ln 2 - k \ln\frac{(2N)!}{(N!)^2} \nonumber \\
&=& O(\ln N)
\end{eqnarray}
In the thermodynamical limit $N\rightarrow \infty$, the mixing
entropy per particle vanishes, and hence the Gibbs paradox is
resolved.

The concept of indistinguishable particles is trivial in Quantum
Mechanics. The identical particles in Quantum Mechanics are
indistinguishable particles. Strictly speaking, one cannot
distinguish two identical particles after their collision. At
equilibrium, the number of microstates of the whole system is
described in terms of the number representation (or the second
quantized representation). The resolution of the Gibbs paradox is
then straightforward in views of the quantum nature of identical
particles.

However, we have also similar case in Classical Mechanics, for
example, colloids. The colloidal particles (giant molecules) of
size up to microns are distinguishable since its behavior should
be governed by classical mechanics. In such a classical system,
the entropy is certainly not extensive \cite{reif,pathria} and the
Gibbs paradox appears. Traditionally, to avoid the Gibbs paradox,
colloidal particles are treated as indistinguishable particles
\cite{chaikin} without any explicit reason. There is still a
puzzle even though this kind of treatment turns out to be correct.

In this paper, we present the thermodynamics of the system of
distinguishable particles, starting from the definition of entropy
for the distinguishable particles. The presentation can easily
show how the ``reduced" entropy \cite{reduced} instead of the
original entropy determines the thermodynamical behaviors.

Before introducing our treatment for distinguishable particles, we
classify the system of particles into three categories,
\begin{enumerate}
    \item Distinguishable particles of the same species
    \item Indistinguishable particles (certainly of the same
    species)
    \item Particles of different species (certainly
    distinguishable)
\end{enumerate}
by their phase spaces. Suppose we call the phase space of the
system of $N$ distinguishable particles of the same species (the
first category) be $\Gamma_N$. The phase space of $N$
indistinguishable particles (the second category) is then
$\Gamma_N / S_N$ with the permutation group $S_N$. The permutation
group $S_N$ is used to eliminate the overcounting microstate
numbers from $\Gamma_N$.

For the third category, each particle belongs to its particular
species. The corresponding phase space of $N$ particles of
different species is
$\Gamma_1\otimes\Gamma_1\otimes\ldots\otimes\Gamma_1$ (total $N$
direct products).

The thermodynamics for the second and third categories were
already well discussed in the textbook \cite{pathria}.

\section{Thermodynamics of distinguishable particles}

To study the thermodynamical variables of the distinguishable
particle system, we consider two subsystems of particle number
$N_1, N_2$, volumes $V_1, V_2$, energies $E_1, E_2$, respectively,
in which their particles, volumes, and energies are allowed to be
exchanged.

Before two subsystems in contact, the phase space is
$\Gamma_{N_1}\otimes\Gamma_{N_2}$. After contact, the phase space
becomes $\Gamma_{N_1+N_2}$ which is larger than
$\Gamma_{N_1}\otimes\Gamma_{N_2}$ by $(N_1+N_2)!/(N_1!N_2!)$
times. This number is the number of way to select $N_1$ particles
and $N_2$ particles from the total $N_1+N_2$ particles. That is,
\begin{eqnarray} \label{phasespace}
\Gamma_{N_1+N_2} =
(\Gamma_{N_1}\otimes\Gamma_{N_2})\oplus(\Gamma_{N_1}\otimes\Gamma_{N_2})\oplus\ldots
\end{eqnarray}
in which there are total $(N_1+N_2)!/(N_1!N_2!)$ copies (or
configurational degeneracy). This degeneracy was pointed out by
Penrose \cite{penrose}.

The total entropy as a whole becomes
\begin{eqnarray} \label{entropy2}
S &=& S_1(N_1,V_1,E_1) + S_2(N_2,V_2,E_2) + k
\ln\frac{(N_1+N_2)!}{N_1!N_2!} \nonumber \\
\end{eqnarray}
$S_1$ and $S_2$ are the entropies of the subsystems separately.
The last term arises from the configurational degeneracy. The
existence of the additional term is due to the nonextensive
property of the system. The distinguishability implies
nonextensivity of the entropy \cite{reif} in which the total
system entropy is not the naive addition of the subsystem
entropies. Similar cases are also well known in other physical
system \cite{nonextensive}.

Because of that, it is easily seen that the Gibbs paradox usually
mentioned in the textbook \cite{pathria} vanishes when we look at
the phase space of the system of distinguishable particles
carefully.

When two subsystems are in equilibrium, the total entropy attains
its maximum, then we have
\begin{eqnarray}
S &=& S_1(N_1,V_1,E_1) + S_2(N_2,V_2,E_2) + S_0(N_1,N_2) \nonumber
\\
\end{eqnarray}
with $S_0 = k\ln((N_1+N_2)!/(N_1!N_2!))$ from Eq.
(\ref{entropy2}). Under the constraint of energy $E_1 + E_2 = E$,
volume $V_1 + V_2 = V$, and particle number $N_1 + N_2 = N$, the
equilibrium attains when the entropy becomes extremum such that
\begin{eqnarray}
\frac{\partial S}{\partial E_1} &=& \frac{\partial S_1}{\partial
E_1} - \frac{\partial S_2}{\partial E_2} = 0 \\
\frac{\partial S}{\partial V_1} &=& \frac{\partial S_1}{\partial
V_1} - \frac{\partial S_2}{\partial V_2} = 0 \\
\frac{\partial S}{\partial N_1} &=& \frac{\partial S_1}{\partial
N_1} - \frac{\partial S_2}{\partial N_2} + \frac{\partial
S_0}{\partial N_1} = 0 \label{chemical}
\end{eqnarray}
We can then obtain, for the system of distinguishable particles,
the expression of the temperature $T$ as
\begin{eqnarray} \label{temperature}
\frac{1}{T} = \left( \frac{\partial S_1}{\partial E_1}
\right)_{V_1,N_1} = \left( \frac{\partial S_2}{\partial E_2}
\right)_{V_2,N_2}
\end{eqnarray}
the pressure $P$ as
\begin{eqnarray} \label{pressure}
\frac{P}{T} = \left( \frac{\partial S_1}{\partial V_1}
\right)_{E_1,N_1} = \left( \frac{\partial S_2}{\partial V_2}
\right)_{E_2,N_2}
\end{eqnarray}
By noticing that
\begin{eqnarray}
\frac{\partial S_0}{\partial N_1} = - \frac{\partial}{\partial
N_1} (k \ln N_1!) + \frac{\partial}{\partial N_2} (k \ln N_2!)
\end{eqnarray}
we can express the chemical potential $\mu$ as
\begin{eqnarray} \label{mu1}
\frac{\mu}{T} = - \left. \frac{\partial ( S_1- k \ln N_1!
)}{\partial N_1} \right|_{E_1,V_1} = - \left. \frac{\partial (
S_2- k \ln N_2! )}{\partial N_2} \right|_{E_2,V_2}  \nonumber \\
\end{eqnarray}
Now we introduce the ``reduced" entropy \cite{reduced}
\begin{eqnarray}
S^{\rm red} = S - k \ln N!
\end{eqnarray}
such that the above expressions for thermodynamical variables
\begin{eqnarray}
\frac{1}{T} &=& \left( \frac{\partial S^{\rm red}}{\partial E}
\right)_{V,N} \\
\frac{P}{T} &=& \left( \frac{\partial S^{\rm red}}{\partial V}
\right)_{E,N} \\
\frac{\mu}{T} &=& - \left( \frac{\partial S^{\rm red}}{\partial N}
\right)_{E,V}
\end{eqnarray}
are re-written in the usual way of thermodynamics for
indistinguishable particles. For the system of distinguishable
particles, the ``reduced" entropy $S^{\rm red}$ instead of the
original entropy $S$ defines the thermodynamical variables of the
system. Our presentation is somehow a straightforward way to
justify the ``reduced" entropy adopted for the classical
(distinguishable) systems in condensed matter physics
\cite{chaikin}.

For the case of indistinguishable particles defined in the phase
space $\Gamma_N/S_N$, its entropy is simply equivalent to the
``reduced" entropy mentioned above. Although the original
entropies for both the distinguishable and indistinguishable
particles are different, the entropies governing their
corresponding thermodynamics are still the same. Thermodynamics
cannot tell the distinguishability of the system.

Suppose our system is $N_1, V_1, E_1$ in contact with the
reservoir of $N_2, V_2, E_2$ in which $N_1 \ll N_2, V_1 \ll V_2,
E_1 \ll E_2$, the entropy of the reservoir can be expanded into
Taylor's series around $N,V,E$ such that
\begin{eqnarray}
S &=& S_1(N_1,V_1,E_1) + S_2(N_2,V_2,E_2) + S_0(N_1,N_2) \nonumber
\\ &=& S_1^{\rm red}(N_1,V_1,E_1) + S_2^{\rm red}(N_2,V_2,E_2) + k
\ln N!  \nonumber
\\ &\simeq& S_2^{\rm red}(N_2,V_2,E_2) + k \ln N!
\end{eqnarray}
in which the system entropy $S_1$ is neglected assuming $S_1 \ll
S_2$. The reduced entropy $S_2^{\rm red}$ can be analyzed by
Taylor's series expansion around $N,V,E$ up to first order, that
is,
\begin{widetext}
\begin{eqnarray}
&&S_2^{\rm red}(N_2,V_2,E_2) \nonumber \\
&=& S_2^{\rm red}(N-N_1,V-V_1,E-E_1)  \nonumber \\
&=& S_2^{\rm red}(N,V,E) -  N_1 \left(\frac{\partial S_2^{\rm
red}}{\partial N_2}\right)_{E,V,N_2=N} - V_1 \left(\frac{\partial
S_2^{\rm red}}{\partial V_2}\right)_{E,N,V_2=V} -
E_1 \left(\frac{\partial S_2^{\rm red}}{\partial E_2}\right)_{E_2=E} \nonumber \\
&=& S_2^{\rm red}(N,V,E) - \frac{1}{T}(-\mu N_1 + E_1 + V_1 P )
\end{eqnarray}
\end{widetext}
 with the temperature of the reservoir $T$.

The total entropy becomes
\begin{eqnarray}
S &=& S_2^{\rm red}(N,V,E) - \frac{1}{T}(-\mu N_1 + E_1 + V_1 P )
+ k \ln N! \nonumber \\
&=& - \frac{1}{T}(-\mu N_1 + E_1 + P V_1 )  + C
\end{eqnarray}
where $C$ is a constant independent of $N_1,E_1,V_1$. The
probability $p(N_1,V_1,E_1)$ of the system is proportional to the
corresponding phase space volume. The phase space we are now
considering is $\Gamma_{N_1+N_2}$, and hence from Eq.
(\ref{phasespace}) there are exactly $N!/(N_1!N_2!)$ copies of the
state characterized by $N_1, V_1, E_1$. We have
\begin{eqnarray}
p(N_1,V_1,E_1) &\propto& \frac{N!}{N_1! N_2!}
\Omega_1(N_1,V_1,E_1)  \nonumber \\
&\propto& \frac{1}{N_1!} \Omega_1(N_1,V_1,E_1) \nonumber \\
&\propto& \frac{1}{N_1!}\exp(\frac{\mu N_1 - E_1 - P V_1}{k T})
\end{eqnarray}
under the condition that $N_1 \ll N_2$.

The probability distribution allows us to formulate the Grand
Canonical ensemble in which our system interacts with a
particle-energy reservoir. The grand partition function
\begin{eqnarray}
\Xi(\mu,V,T) &=& \sum_{N_r = 0}^\infty \sum_{s} \frac{1}{N_r!}
{\rm e}^{\mu N_r/k T }  {\rm e}^{- E_s/k T}  \nonumber \\
&=& \sum_{N_r = 0}^\infty {\rm e}^{\mu N_r/k T} {\cal
Z}_{N_r}(V,T)
\end{eqnarray}
with the partition function
\begin{eqnarray}
{\cal Z}_{N_r}(V,T)= \frac{1}{N_r!} \sum_{s} {\rm e}^{- E_s/k T}
\end{eqnarray}
of the corrected Boltzmann counting due to the factor $1/N_r!$.

\section{Conclusion}

In summary, although the phase spaces of the system of
distinguishable particles is different from that of
indistinguishable one, their thermodynamics are in fact
equivalent. It also implies that the corrected Boltzmann counting
factor can be justified in classical statistical mechanics.

\section{Acknowledgements}

The author thanks Prof. Pik-Yin Lai and Prof. Shin-Chuan Gou for
their helpful comments and discussions. Very useful comment from
the reviewer is highly acknowledged. Support from the National
Science Council of the Republic of China is acknowledged.


\begin{thebibliography}{}
\bibitem{reduced}
The ``reduced entropy" is $S-k\ln N!$ in which $S$ is the original
entropy and $N$ is the number of particles in the system. We
adopted the terminology because it is the same as the entropy of
the system assuming that the particles in the system are
indistinguishable. ``Reduced" here means the reduction of the
distinguishablility degrees of freedom.
\bibitem{nonextensive}
For example, if we consider the surface energy of the system, then
the energy of the system is not proportional to its mass
(nonextensive). When two subsystems are in contact, their total
energy are not equal to the sum of the original two energies.
Another example is the system with long-ranged electrostatic
interaction.
\end{thebibliography}
\end{document}